\documentclass[a4paper,12pt]{article}
\usepackage{mathrsfs}
\usepackage{graphicx} 
\usepackage{epstopdf}

\begin{document}

\hoffset = -1truecm \voffset = -2truecm \baselineskip = 10 mm

\title{\bf Improved Bethe-Heitler formula}

\author{
{\bf Wei Zhu} \\
\normalsize Department of Physics, East China Normal University,
Shanghai 200241, China\\
}

\date{}

\newpage

\maketitle
\begin{abstract}

       The Bethe-Heitler formula describes
bremsstrahlung and it's a typical and important example in quantum
electromagnetic dynamics (QED). This formula is widely applied in
many branches of physics and astrophysics. We find that the
integrated bremsstrahlung cross section at the static approximation
and high energy limit has an unexpected big increment, which is
missed by previous bremsstrahlung theory. This anomalous effect also
exists in electron-positron pair creation. We derive the relating
formulas and point out that electromagnetic cascades at the top of
atmosphere can test this effect.

\end{abstract}

{\bf keywords}:  Anomalous effect; Bremsstrahlung; Pair creation;
Electromagnetic cascades

\vskip 1truecm

\newpage
\begin{center}
\section{\bf Introduction}
\end{center}

    When electrons scatter off electric field of proton or nucleus,
they can emit real photons. This is bremsstrahlung (braking
radiation). Bethe and Heitler first gave a quantum description of
the bremsstrahlung emission with the Coulomb potential of an
infinite heavy atom [1]. The Bethe-Heitler (BH) formula is an
elementary and important equation in quantum electromagnetic
dynamics (QED) and astrophysics.

    Recently, a puzzled difference of the energy spectra of electrons and
positrons at GeV-TeV energy band in cosmic rays raises our doubts to
the validity of the BH formula at high energy. These energy spectra
have been measured at the atmosphere top by Alpha Magnetic
Spectrometer (AMS)[2], Fermi Large Area Telescope (Fermi-LAT)[3],
DArk Matter Particle Explorer (DAMPE)[4] and Calorimetric Electron
Telescope (CALET)[5]. The discovery of the excess (or break) of the
spectra in the GeV-TeV band causes big interest because it may be
related to the new physical signals including dark matter. However,
a puzzled question is why the data of AMS and CALET are noticeably
lower than that of DAMPE and Fermi-LAT at the measured energy band?
This uncertainty in the key range makes us unable to understand the
meaning of the signal correctly. The above measurements have been
accumulated and improved over many years. Besides, Fermi-LAT, DAMPE
and CALET use the similar calorimeter, while AMS employs a
completely different kind of magnetic spectrometer. Therefore, the
above difference seems not to be caused by the systematic or
measurement errors.

     We noticed that both AMS and CALET set on the international
space station at $\sim 400~km$, while Fermi-LAT and DAMPE are
orbiting the Earth at $500\sim560~km$ altitude. A naive suggestion
is that the primary signals of electron-positron fluxes are weakened
by the electromagnetic shower caused by the extremely thin
atmosphere during its transmission from $500~km$ to $400~km$. For
this sake, we use the electromagnetic cascade equation to estimate
the value of the corresponding radiation length $\lambda$, which may
lead to the difference between the spectra as shown in Fig. 1. We
find that $\lambda\simeq 10^{-6} g/cm^2$. This value is seven-orders
of magnitude smaller than the usual standard value
$\lambda=37~g/cm^2$.

\begin{figure}
    \begin{center}

 \includegraphics[width=0.8\textwidth]{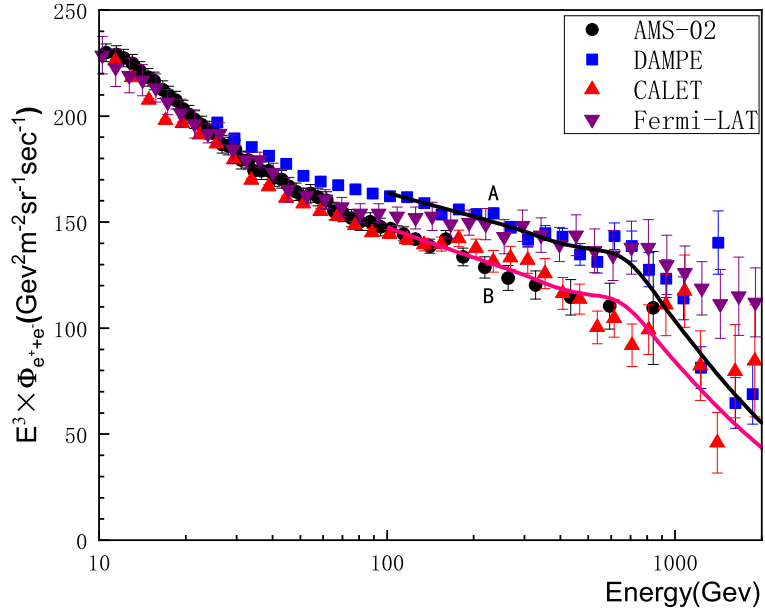}
        \caption{Cosmic electron-positron spectra multiplied by $E^3$ as a
function of energy. Data are taken from [2-5]. A and B indicate the
spectra at height $\sim 500$ and $\sim 400$ km.  The curve A is an
input of the electromagnetical cascade equation, and the curve B is
the result of the cascade through $\sim 0.1\lambda$. The
corresponding bremsstrahlung cross section is seven-orders of
magnitude larger than the prediction of the traditional theory. The
difference between the curves A and B is explained as an anomalous
bremsstrahlung effect in this work.}\label{Fig1}

    \end{center}
\end{figure}

    Is it possible that there exists such a big difference in
the radiation length? In order to explore the possibility of
bremsstrahlung enhancement, we review the derivation of the BH
formula. The bremsstrahlung event contains the scattering of the
incident electron on the nuclear electric field for the conservation
of energy-momentum. It is well known that the total cross section of
the Rutherford cross section in a pure Coulomb field either in the
classical or quantum theory is infinite, which origins from the
following fact: the long-range $1/r$ potential has a significant
contribution to the total cross section. If the impact parameter of
the incident electron is large compared with the atomic radius, the
Coulomb field of the positive nuclear charge is completely screened
by the electrons of a neutral atom. Therefore, the Coulomb cross
section is limited in an atomic scale $R^2\sim 1/\mu^2$, $\mu$ is
the screening parameter (Fig. 2a). On the other hand, Bethe and
Heitler predict a strongly reduced bremsstrahlung cross section
$\sim 1/m^2_e$, which is much smaller than the geometric scattering
cross section (Fig. 2b).

\begin{figure}
    \begin{center}

        \includegraphics[width=0.8\textwidth]{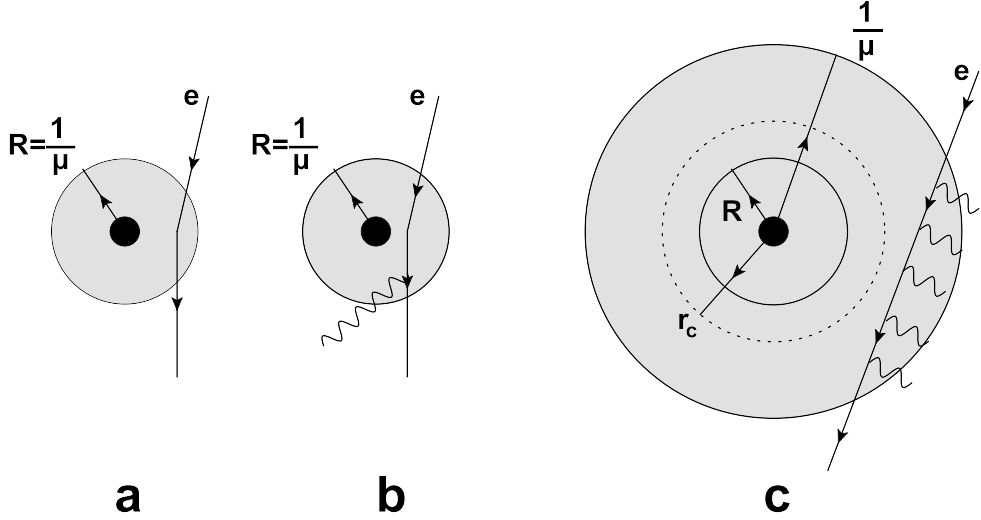}
\caption{(a) The Rutherford cross section $\sim Z^2\alpha^2/\mu^2$;
(b) The Bethe-Heitler cross section $\sim
Z^2\alpha^3/m^2_e\ln(m^2_e/\mu^2)$; (c) The anomalous bremsstrahlung
cross section $\sim Z^{*2}\alpha^3/\mu^2\ln(4E^2_i/\mu^2)$, $Z^*$ is
the affective ionized charge number.} \label{Fig2}

    \end{center}
\end{figure}

    In order to find the reason of the difference presented in Fig. 2,
we expose how a factor $1/m^2_e$ replaces $1/\mu^2$ in the BH
formula. Unfortunately, the complex correlations between the
scattering and radiation processes hindered Bethe and Heitler to
obtain an analytical solution for the integrated cross section if
the screening parameter was considered at the beginning derivation
in the $S$-matrix. In deed, the parameter $\mu$ is introduced by a
model at the last step in their derivation. Therefore, we can't
track the whereabouts of the screening parameter in such way.

     We re-derive the bremsstrahlung formula with the screening
potential at high energy. We find that if considering energy
transfer due to the recoil effect, the interference amplitudes can
be removed if the energy of electron is high enough. Thus we can
further decompose the process, where the time ordered perturbative
theory (TOPT) [6,7] is used to separate the scattering and radiation
processes at the equivalent photon (Weizs\"{a}cker-Williams)
approximation [8,9]. The above simplified method allows us to track
how the screening parameter enters the final cross section from an
original $S$-matrix element. Using this method, we find that the
parameter $m_e$ in the radiation part enter into the denominator of
the scattering part through a simple mathematical formula.

    Let us use a simple example to illustrate our discovery.
A typical integrated Rutherford cross section contains

$$d\sigma_{Ruth}\sim \int^{\pi}_0\frac{\sin
\theta d\theta}{(\sin^2\frac{\theta}{2}+\frac{\mu^2}{4E_i^2})^2}\sim
\frac{1}{\mu^2},\eqno(1.1)$$where $\theta$ is the scattering angle
and $E_i$ the initial energy of electron. The result is proportional
to the geometric area of an atomic electric field. Although
scattering away from the target is weak, the cumulative
contributions of scattering in a broad space lead to the divergence
of the total cross section at $\mu\rightarrow 0$. On the other hand,
a radiation factor combines with the scattering matrix element in
bremsstrahlung and the integrated cross section becomes

$$d\sigma_{Bress}\sim \int^{\pi}_{\theta_{min}}\frac{\sin \theta
d\theta}{(\sin^2\frac{\theta}{2}+\frac{\mu^2}{4E_i^2})^2}\ln\frac{-q^2}{B}\sim
\frac{1}{\mu^2+m^2_e}\sim \frac{1}{m^2_e},\eqno(1.2)$$where a weakly
$\mu$-dependent logarithm is neglected. Note that
$-q^2=m^2_e+4E^2_i\sin^2(\theta/2)$ and $B=m^2_e$ or $\mu^2$. One
can clearly see that the lower limit $\theta_{min}$ of the
integration is determined by the condition $-q^2\geq B$. Once there
is a un-eliminated parameter $m_e$ in $-q^2/B$, it will enter into
the denominator after integrating the scattering angle. The result
indicates that the bremsstrahlung cross section has a strong
suppression since $\mu\ll m_e$.

    A following interesting question is under what condition the parameter $m_e$
can be omitted in $\ln (-q^2/B)$? In this case, the contributions of
the geometric cross section $1/\mu^2$ will be restored. The
straightforward answer is

$$\ln\frac{-q^2}{\mu^2}\rightarrow
\ln\frac{\vert\vec{q}\vert^2}{\mu^2}, \eqno(1.3)$$where
$\vec{q}^2=4E^2_i\sin^2(\theta/2)$ (at $v\rightarrow c$). It
means a no-recoil scattering. In this case, the bremsstrahlung cross
section restores its geometric size and we call this as the
anomalous bremsstrahlung effect. Theoretically, an infinite heavy
atom can completely absorb the recoil effect as the Rutherford
scattering. However, a target atom bound in the normal mater can not
avoid the obviously recoil corrections due to the strong collisions.
Therefore, the suppression in the bremsstrahlung cross section is a
general phenomenon.

    However, there is an exception as we have mentioned at the
beginning of this work. In the complete ionosphere about $400\sim
500~km$ height, the oxygen atoms are not only completely ionized,
but its density is extremely thin. On average, there is only one
atom per 1/1000000000 cubic centimeter. This is a big space with
macroscopic scale $\sim 10^{-3}~cm$. The nuclear Coulomb potential
may spread into such a broader space, where the bremsstrahlung
events may neglect the recoil energy comparing with a larger initial
energy $E_i$ of electron, since they are far away from the source of
the Coulomb center field. Besides, the integration on the space may
go down to a lower limit $\vert\vec{q}\vert^2=\mu^2$ no blocking
from the cut-parameter $m^2_e$. There is a critical scale $r_c$,
when the impact parameter larger than $r_c$ (Figure 2c), where the
bremsstrahlung cross section will restore the big geometric cross
section. Thus, one can get a large enough increment of the cross
section to explain the result in Fig. 1 since the accumulation of a
large amount of soft photon radiation in a broad space.

   We emphasize that $m_e$ and $\mu$ have different physical meaning
although they both have the mass dimension in the natural units.
Therefore, when $m_e$ in $\ln(-q^2/\mu^2)$ is omitted due to the
recoil energy $\nu\ll E_i$, a more smaller parameter $\mu$ can be
retained because $\mu$ is irrelevant to the energy $E_i$ of the
incident electron.

    According to the above considerations, we derived a new bremsstrahlung formula, they are

$$d\sigma^{I}=\frac{\alpha^3}{m_e^2}\ln\frac{4E_i^2}{\mu^2}
\frac{[1+(1-z)^2](1-z)}{z}dz\rightarrow
\frac{2\alpha^3}{m_e^2}\ln\frac{4E_i^2}{\mu^2}
\frac{dz}{z}\eqno(1.4)
$$in the normal media and

$$d\sigma^{II}=\frac{2\alpha^3}{\mu^2}\ln\frac{4E_i^2}{\mu^2}
\frac{[1+(1-z)^2](1-z)}{z}dz\rightarrow
\frac{4\alpha^3}{\mu^2}\ln\frac{4E_i^2}{\mu^2}
\frac{dz}{z}\eqno(1.5)$$ in the thin ionized gas, where
$z\simeq\omega/E_i$ and taking the leading logarithmic $(1/z)$
approximation. We will prove that $d\sigma^I$ is compatible with the
BH formula.

    Our discussion about the bremsstrahlung process also applies to
pair production of electron-positron. We derived the improved
formula. For testing the above anomalous effect, a modified cascade
equation for the electromagnetic shower is given.

    The paper is organized as follows. In Sec. 2 we detail the
derivation of the bremsstrahlung formula using the TOPT. Then we discuss
the BH formula and compare these different versions for the
bremsstrahlung formula at Sec.
3. The anomalous effect in electron-positron
pair creation is studied at Sec. 4. The improved cascade equation
for electromagnetic shower is given at Sec. 5. The last section is a
summary.

\newpage
\begin{center}
\section{\bf The bremsstrahlung cross section with screening potential}
\end{center}

    The BH formula assumes that the target atom is
infinitely heavy. We consider a more general case in the following
derivation: electron scattering off a finite heavy atom. The
differential cross section of the bremsstrahlung emission (Fig. 3)
in covariant perturbation theory at the leading order approximation
is [10]

$$d\sigma=\frac{m_eM_0}{\sqrt{(p_iP_i)^2-m^2_eM^2_0}}\vert\overline{M_{p_iP_i\rightarrow
p_fP_fk}}\vert^2(2\pi)^4\delta^4(p_i+P_i-p_f-P_f-k)$$
$$\frac{2\pi d^3\vec{k}}{(2\pi)^3\omega}\frac{m_ed^3\vec{p_f}}{(2\pi)^3E_f}
\frac{M_0d^3\vec{P_f}}{(2\pi)^3E^P_f},\eqno(2.1)$$where the
screening photon propagator in the matrix takes

$$\frac{ig_{\mu\nu}}{q^2-\mu^2+i\epsilon}.\eqno(2.2)$$
Its 3-dimension component $1/(\vec{q}^2+\mu^2)$ corresponding to a
potential $\sim e^{-r\mu}/r$, i.e. the Coulomb potential vanishes at
$r>1/\mu\equiv R$. $R$ is the atom radius for a neutral atom.

    According to the TOPT, a covariant
Feynman propagator in $M_{p_iP_i\rightarrow p_fP_fk}$

$$S=\int d^4l\frac{i\gamma\cdot
l+m_e}{l^2-m^2_e+i\epsilon},\eqno(2.3)$$may decompose into a forward
and a backward components (Fig. 4)

$$S_F=\frac{1}{2E_{\hat{l}}}\frac{i\gamma\cdot\hat{l}+m_e}{\omega+E_f-E_{\hat{l}}},~~forward~~\eqno(2.4)$$and

$$S_B=\frac{1}{2E_{\hat{l}}}\frac{i\gamma\cdot\hat{l}+m_e}{-\omega+E_f-E_{\hat{l}}}.~~backward~~\eqno(2.5)$$
Note that $l(E_l,\vec{l}_T,l_L)$ is off-mass shell $l^2\neq m^2_e$,
while $\hat{l}=(E_{\hat{l}}, \vec{l}_T, l_L)$ or
$\hat{l}=(-E_{\hat{l}}, \vec{l}_T, l_L)$ are on-mass shell, i.e.,
$\hat{l}^2=m^2_e$.

\begin{figure}
    \begin{center}
        \includegraphics[width=0.6\textwidth]{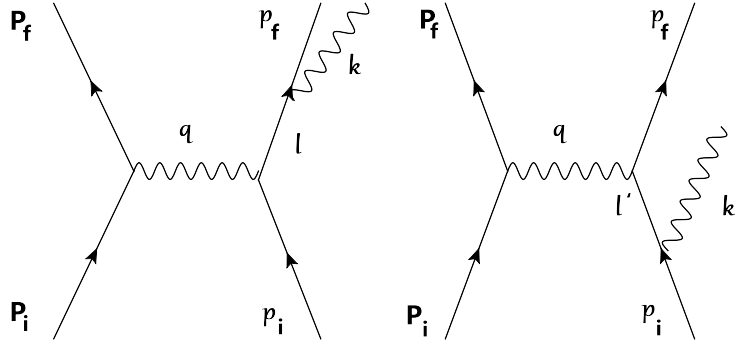}
        \caption{Two elemental bremsstrahlung amplitudes.}\label{Fig3}
    \end{center}
\end{figure}

\begin{figure}
    \begin{center}
        \includegraphics[width=0.5\textwidth]{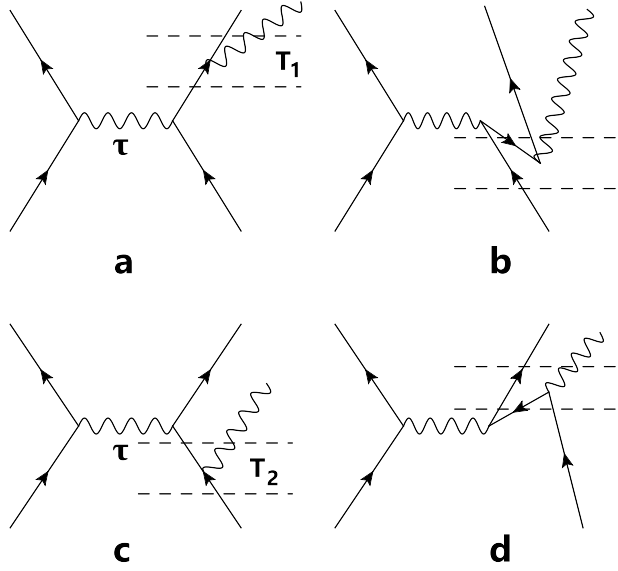}
        \caption{The TOPT decomposition of Fig.3. The dashed lines
            indicate the time ordered of the process.}\label{Fig4}
    \end{center}
\end{figure}

    The physical picture is frame-dependent in the TOPT and it is not a
relativistically invariant. The same physical process has different
appearances in different frames, simple in some but complicated in
others [11]. It seems that the TOPT decomposition complicates the
calculation with increasing the propagators. However, the backward
component will be suppressed at higher energy and small emitted
angle. For example, we take $\vec{\hat{l}}$ along the z-direction,
and define $z$ as the momentum fraction of $\hat{l}$ carried by the
longitudinal momentum $k_L$ of photon,

$$z=\frac{k_L}{\hat{l}}\simeq
\frac{\omega}{E_{\hat{l}}},\eqno(2.6)$$where we neglect the electron
mass $m_e$ at high energy and note that $\hat{l}$ is on-mass shell.
At high energy and small emitted angle we denote

$$\hat{l}=(E_{\hat{l}}, \hat{l}_T, \hat{l}_L)=(E_{\hat{l}}, 0,
E_{\hat{l}}), \eqno(2.7)$$

$$k=(\omega, k_T, k_L)=\left(zE_{\hat{l}}+\frac{k_T^2}{2zE_{\hat{l}}}, k_T, zE_{\hat{l}}\right),
\eqno(2.8)$$and

$$p_f=(E_f, p_{f,T},
p_{f,L})=\left((1-z)E_{\hat{l}}+\frac{k^2_T}{2(1-z)E_{\hat{l}}},
-k_T, (1-z)E_{\hat{l}}\right). \eqno(2.9)$$  If $z\neq 0$,
$1$ and $E_i\gg k_T$ one can find that

$$S_F\simeq\frac{1}{2E_{\hat{l}}}\frac{1}{\omega+E_f-E_{\hat{l}}}\simeq
\frac{z(1-z)}{k^2_T}, \eqno(2.10)$$which is much larger than

$$S_B\simeq\frac{1}{2E_{\hat{l}}}\frac{1}{\omega-E_f+E_{\hat{l}}}\simeq
\frac{1}{4zE^2_{\hat{l}}}. \eqno(2.11)$$Therefore, the contributions
of the backward propagator are negligible at high energy. This not
only reduces the number of diagrams, but also allows us to factorize
the complex Feynman graph due to the on-mass shell of the forward
propagator. This is the theoretical base of the equivalent photon
approximation.

\begin{figure}
    \begin{center}
        \includegraphics[width=0.6\textwidth]{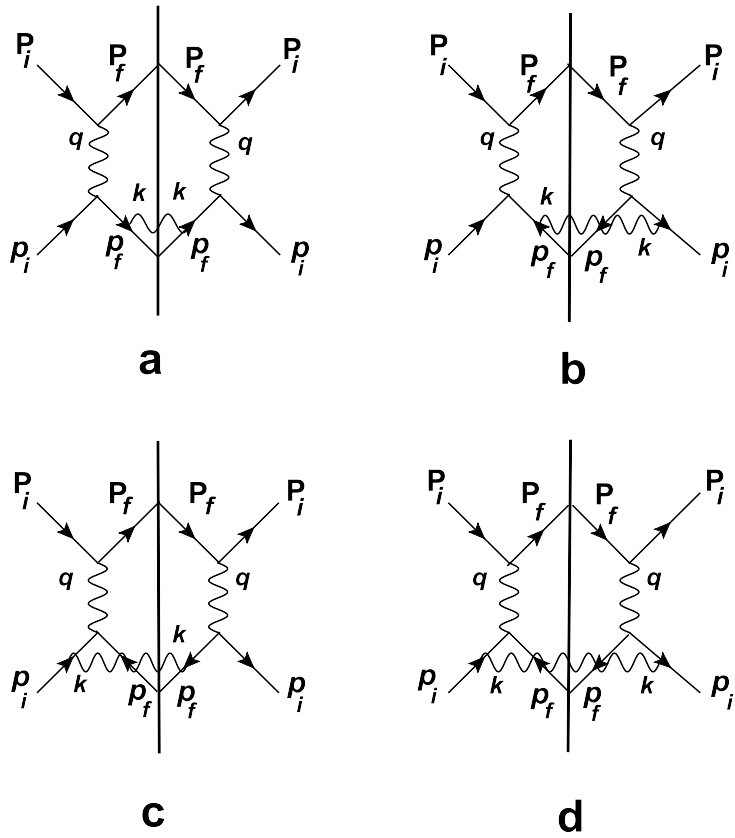}
        \caption{Four TOPT diagrams after neglecting the
            contributions of the backward components at high energy and small
            scattering angle.}\label{Fig5}
    \end{center}
\end{figure}

    We take the laboratory frame, where
the target atom is at rest, but the incident electron has a high
energy. This is an infinite momentum frame for the electron. In the
above mentioned laboratory frame, both the longitudinal momentum and
energy of the virtual photon cannot be ignored. Thus, the
contributions of Figs. 5b and 5c are not negligible due to the
coherence between Fig. 4a and 4c. The electron-atom scattering time
is

    $$\tau \sim \frac{1}{\nu},\eqno(2.12)$$ $\nu$ is the energy of the virtual photon.
The radiation time is

$$T_{1,2}\sim \frac{1}{E_f+\omega-E_{\hat{l}}}=\frac{E_{\hat{l}}z(1-z)}{k^2_T}, \eqno(2.13)$$
during this period the photon is emitted. Since $E_{\hat{l}}\propto
E_i$ and $E_{\hat{l}}\gg k_T$, at high enough energy $E_i$ and
$z\neq 0,1$ we always have

$$\tau < T_{1,2}\eqno(2.14)$$for a not too large value of $\nu$.

    The virtual photon $\gamma^*$ with a short life $\tau$ triggers the
event Fig. 4c (or the event Fig. 4a) needs time $T_2$ (or $T_1$).
Note that Fig. 4a,4c are time-ordered processes in the TOPT
description since the backward components are suppressed. This
virtual photon can't trigger the following event 4a if it has
triggered the event 4c. It also can't trigger the event 4c before it
triggers the event 4a. It implies that the processes Fig. 4a,4c are
incoherent. Therefore, the contributions of the interferant
processes in Fig. 5b,5c can be neglected in our following
discussions. After removing these coherent diagrams, using the
on-mass shell of the momentum $\hat{l}$, the process can further
decompose into two sub-processes.

    We discuss the process in Fig. 5a, Eq. (2.1) becomes

$$d\sigma_a^I=\frac{m_eM_0}{\sqrt{(p_iP_i)^2-m^2_eM^2_0}}\vert\overline{M_{p_iP_i\rightarrow
\hat{l}P_f}}\vert^2\frac{M_0d^3\vec{P_f}}{(2\pi)^3E^P_f}\frac{m_ed^3\vec{p_f}}{(2\pi)^3E_f}
(2\pi)^4\delta^4(p_i+P_i-p_f-P_f-k)$$
$$\left(\frac{1}{2E_{\hat{l}}}\right)^2\left(\frac{1}{E_f+\omega-E_{\hat{l}}}\right)^2
\vert\overline{M_{\hat{l}\rightarrow p_fk}}\vert^2 \frac{2\pi
d^3\vec{k}}{(2\pi)^3\omega}$$
$$\equiv d\hat{\sigma}_ad{\cal{P}}_a ,\eqno(2.15)$$where

$$d\hat{\sigma}_a^I=\frac{m_eM_0}{\sqrt{(p_iP_i)^2-m^2_eM^2_0}}\vert\overline{M_{p_iP_i\rightarrow
\hat{l}P_f}}\vert^2\frac{M_0d^3\vec{P_f}}{(2\pi)^3E^P_f}\frac{m_ed^3\vec{p_f}}{(2\pi)^3E_f}
(2\pi)^4\delta^4(p_i+P_i-p_f-P_f-k),\eqno(2.16)$$and

$$d{\cal{P}}_a=\frac{1}{4\pi^2}\left(\frac{1}{2E_{\hat{l}}}\right)^2\left(\frac{1}{E_f+\omega-E_{\hat{l}}}\right)^2
\vert\overline{M_{\hat{l}\rightarrow p_fk}}\vert^2
\frac{d^3\vec{k}}{\omega}. \eqno(2.17)$$

    We calculate $d\hat{\sigma}_a^I$ using

$$\vert\overline{M_{p_iP_i\rightarrow\hat{l}P_f}}\vert^2
=\frac{e^2(Ze)^2(4\pi)^2}{4m_e^2M^2_0(q^2-\mu^2)^2}
\left[l^{\mu}p^{\nu}_i+p^{\mu}_il^{\nu}-g^{\mu\nu}(l\cdot
p_i-m^2_e)\right]$$
$$\left[P^{\mu}_fP^{\nu}_i+P^{\mu}_iP^{\nu}_f-g^{\mu\nu}(P_f\cdot
P_i-M^2_0)\right], \eqno(2.18)$$where we use $l$ to replace
$\hat{l}$ in the matrix since $E_l\simeq E_{\hat{l}}$ for the small
emitted angle. The result is

$$d\hat{\sigma}_a^I=\frac{Z^2\alpha^2}{4 E_i^2}\frac{1}{
(\sin^2\frac{\theta}{2}+\mu^2/4E_i(E_f+\omega))^2}
\frac{\cos^2\frac{\theta}{2}-\frac{q^2}{2M^2_0}\sin^2\frac{\theta}{2}}
{1+\frac{2E_i}{M_0}\sin^2\frac{\theta}{2}}d\Omega$$
$$=\frac{Z^2\alpha^2}{4 E_i^2}\frac{1}{
((1+\frac{\mu^2}{2E_iM_0})\sin^2\frac{\theta}{2}+\mu^2/4E_i^2)^2}
\frac{\cos^2\frac{\theta}{2}-\frac{q^2}{2M^2_0}\sin^2\frac{\theta}{2}}
{1+\frac{2E_i}{M_0}\sin^2\frac{\theta}{2}}d\Omega$$
$$\simeq\frac{Z^2\alpha^2}{4 E_i^2}\frac{1}{
(\sin^2\frac{\theta}{2}+\mu^2/4E_i^2)^2}
\frac{\cos^2\frac{\theta}{2}-\frac{q^2}{2M^2_0}\sin^2\frac{\theta}{2}}
{1+\frac{2E_i}{M_0}\sin^2\frac{\theta}{2}}d\Omega. \eqno(2.19)$$
Note that the $q^2$-dependent term in Eq. (2,19) is absent when the
target is a spin-0 particle, however, it does not change the
following results.

    A key point is that the TOPT allows us to separately calculate the splitting
function ${\cal{P}}_a$ and the sub-cross section $d\hat{\sigma}$.
For the sub-process $e(\hat{l})\rightarrow e(p_f)+\gamma(k)$, we let
$z$-axis alon the direction of $\vec{\hat{l}}$. Using Eqs.
(2.7)-(2.9) we have

$$\frac{1}{E_f+\omega-E_{\hat{l}}}=\frac{2E_{\hat{l}}z(1-z)}{k^2_T},\eqno(2.20)$$

$$\frac{d^3\vec{k}}{(2\pi)^32\omega}\simeq
\frac{z(1-z)dk^2_T}{16\pi^2z},\eqno(2.21)$$and

$$\vert\overline{M_{\hat{l}\rightarrow
p_fk}}\vert^2=2\alpha\frac{1+(1-z)^2}{z^2(1-z)}k^2_T,\eqno(2.22)$$we
obtain

$$d{\cal{P}}_a=\frac{\alpha}{2\pi}\frac{[1+(1-z)^2](1-z)}{z}dzd\ln
k^2_T. \eqno(2.23)$$The calculation is accurate to $\ln k^2_T$ since
$\vert \vec{k}_T\vert \ll E_{\hat{l}}$.

    Combining equations (2.19) and (2.23), we have

$$d\sigma_a^I=\frac{Z^2\alpha^2}{4 E_i^2}\frac{1}{
(\sin^2\frac{\theta}{2}+\mu^2/4E_i^2)^2}
\frac{\cos^2\frac{\theta}{2}-\frac{q^2}{2M^2_0}\sin^2\frac{\theta}{2}}
{1+\frac{2E_i}{M_0}\sin^2\frac{\theta}{2}}
\frac{\alpha}{2\pi}\frac{[1+(1-z)^2](1-z)}{z}d\Omega dzd\ln k^2_T$$
$$=\frac{Z^2\alpha^2}{4 E_i^2}\frac{1}{
(\sin^2\frac{\theta}{2}+\mu^2/4E_i^2)^2}
\frac{1-\sin^2\frac{\theta}{2}+\frac{2E^2_i}{M^2_0}\frac{\sin^4\frac{\theta}{2}}
{1+\frac{2E_i}{M_0}\sin^2\frac{\theta}{2}}}
{1+\frac{2E_i}{M_0}\sin^2\frac{\theta}{2}}
\frac{\alpha}{2\pi}\frac{[1+(1-z)^2](1-z)}{z}d\Omega dzd\ln k^2_T.
\eqno(2.24)$$

    For a virtual mass $-q^2$, the integral in $k^2_T$ has an upper
limit of order $-q^2$ since $k_T^2$ origins from $q^2$. Thus, we
have

    $$\int_{k_{T,min}}^{k_{T,max}}\frac{d k^2_T}{k^2_T}=
\ln\frac{-q^2}{\mu^2}, \eqno(2.25)$$where we have $-q^2>\mu^2$ since
$k^2_T$ is ordered and $\mu$ is a minimum quantity with the mass
dimension. Thus, we can sum maximum contributions to the cross
section.

    The energy momentum conservation

$$\nu=E_i-E_f-\omega=\frac{2E_i(E_f+\omega)}{M_0}\sin^2\frac{\theta}{2},
\eqno(2.26)$$and

$$E_f+\omega=E_i\left(1+\frac{2E_i}{M_0}\sin^2\frac{\theta}{2}\right)^{-1}.\eqno(2.27)$$
Note that the 4-transfer momentum

$$q^2=2m^2_e-4E_i(E_f+\omega)\sin^2\frac{\theta}{2}\simeq
2m^2_e-4E_i^2\sin^2\frac{\theta}{2},  \eqno(2.28)$$where we take
$1+2E_i/M_0\sin^2(\theta/2)\simeq 1$ since the leading contributions
are from $\theta\rightarrow \theta_{min}$.

   We calculate the integrated bremsstrahlung cross section at a given initial energy through the angle-integral.
In general the term $2m^2_e$ in Eq. (2.28) can not be omitted since
the value of $-q^2$ is not always large even at high energy. The
lower limit $\theta_{min}$ of integral is determined by

$$\ln\frac{-2m^2_e+4E^2_i\sin^2\frac{\theta}{2}}{\mu^2}=
\ln\frac{4E_i^2t'}{\mu^2}=\ln\frac{4E_i^2}{\mu^2}+\ln t'\geq
0,\eqno(2.29)$$where $\sin^2(\theta/2)\equiv t$ and $t'\equiv
t-m^2_e/(2E^2_i)$.

    Substituting Eqs. (2.25) and (2,28) into Eq. (2.24), we divide the integral into
three parts.

$$d\sigma_a^I(1)=\frac{Z^2\alpha^2}{4
E_i^2}\ln\frac{4E_i^2}{\mu^2}2\pi\int^{\pi}_{\theta_{min}}\frac{sin
\theta d\theta}{(\sin^2\frac{\theta}{2}+\mu^2/4E_i^2)^2} \frac{1}
{1+\frac{2E_i}{M_0}\sin^2\frac{\theta}{2}}
\frac{\alpha}{2\pi}\frac{[1+(1-z)^2](1-z)}{z}dz$$
$$=\frac{\pi Z^2\alpha^2}{2E_i^2}\ln\frac{4E_i^2}{\mu^2}\frac{M_0}{2E_i}\int^{\pi}_{\theta_{min}}\frac{2sin
\frac{\theta}{2}cos\frac{\theta}{2}
d\theta}{(\sin^2\frac{\theta}{2}+\mu^2/4E_i^2)^2} \frac{1}
{\frac{M_0}{2E_i}+sin^2\frac{\theta}{2}}
\frac{\alpha}{2\pi}\frac{[1+(1-z)^2](1-z)}{z}dz$$
$$=\frac{\pi Z^2\alpha^2}{2E_i^2}\ln\frac{4E_i^2}{\mu^2}\frac{M_0}{2E_i}\int^1_{\epsilon}
\frac{2dt}{(t+\mu^2/4E_i^2)^2} \frac{1}{\frac{M_0}{2E_i}+t}
\frac{\alpha}{2\pi}\frac{[1+(1-z)^2](1-z)}{z}dz$$
$$=\frac{\pi Z^2\alpha^2}{2E_i^2}\ln\frac{4E_i^2}{\mu^2}\frac{M_0}{E_i}
\frac{1}{\frac{\mu^2}{4E_i^2}-\frac{M_0}{2E_i}}
\left(\frac{1}{1+\frac{\mu^2}{4E_i^2}}
+\frac{1}{\frac{M_0}{2E_i}-\frac{\mu^2}{4E^2_i}}\ln\frac{1+\frac{\mu^2}{4E^2_i}}{1+\frac{M_0}{2E_i}}
-\frac{1}{\epsilon+\frac{\mu^2}{4E^2_i}}-\frac{1}{\frac{M_0}{2E_i}+\frac{\mu^2}{4E^2_i}}
\ln\frac{\epsilon+\frac{\mu^2}{4E^2_i}}{\epsilon+\frac{M_0}{2E_i}}\right)$$
$$\times\frac{\alpha}{2\pi}\frac{[1+(1-z)^2](1-z)}{z}dz$$
$$\simeq
\frac{Z^2\alpha^3}{m^2_e}\ln\frac{4E^2_i}{\mu^2}\frac{[1+(1-z)^2](1-z)}{z}dz,\eqno(2.30)$$where
$t=sin^2(\theta/2)$ and $\epsilon=t_{min}=m^2_e/(2E^2_i)$. One can
find that $m_e^2$ comes from the lower limit $\epsilon$.

$$d\sigma_a^I(2)=-\frac{Z^2\alpha^2}{4
E_i^2}\ln\frac{4E_i^2}{\mu^2}2\pi\int^{\pi}_{\theta_{min}}\frac{sin
\theta d\theta}{(\sin^2\frac{\theta}{2}+\mu^2/4E_i^2)^2}
\frac{sin^2\frac{\theta}{2}}
{1+\frac{2E_i}{M_0}\sin^2\frac{\theta}{2}}
\frac{\alpha}{2\pi}\frac{[1+(1-z)^2](1-z)}{z}dz$$
$$=-\frac{\pi Z^2\alpha^2}{2E_i^2}\ln\frac{4E_i^2}{\mu^2}\frac{M_0}{2E_i}\int^{\pi}_{\theta_{min}}\frac{2sin
\frac{\theta}{2}cos\frac{\theta}{2}
d\theta}{(\sin^2\frac{\theta}{2}+\mu^2/4E_i^2)^2}
\frac{sin^2\frac{\theta}{2}}
{\frac{M_0}{2E_i}+sin^2\frac{\theta}{2}}
\frac{\alpha}{2\pi}\frac{[1+(1-z)^2](1-z)}{z}dz$$
$$=-\frac{\pi Z^2\alpha^2}{2E_i^2}\ln\frac{4E_i^2}{\mu^2}\frac{M_0}{2E_i}\int^1_{\epsilon}
\frac{2dt}{(t+\mu^2/4E_i^2)^2} \frac{t}{\frac{M_0}{2E_i}+t}
\frac{\alpha}{2\pi}\frac{[1+(1-z)^2](1-z)}{z}dz$$
$$=-\frac{\pi Z^2\alpha^2}{2E_i^2}\ln\frac{4E_i^2}{\mu^2}\frac{M_0}{E_i}
\frac{1}{\frac{\mu^2}{4E_i^2}-\frac{M_0}{2E_i}}
\left(\frac{\frac{\mu^2}{4E^2_i}}{1+\frac{\mu^2}{4E_i^2}}
+\frac{\frac{M_0}{2E_i}}{\frac{M_0}{2E_i}-\frac{\mu^2}{4E^2_i}}\ln\frac{1+\frac{\mu^2}{4E^2_i}}{1+\frac{M_0}{2E_i}}
-\frac{\frac{\mu^2}{4E^2_i}}{\epsilon+\frac{\mu^2}{4E^2_i}}-\frac{\frac{M_0}{2E_i}}{\frac{M_0}{2E_i}+\frac{\mu^2}{4E^2_i}}
\ln\frac{\epsilon+\frac{\mu^2}{4E^2_i}}{\epsilon+\frac{M_0}{2E_i}}\right)$$
$$\times\frac{\alpha}{2\pi}\frac{[1+(1-z)^2](1-z)}{z}dz. \eqno(2.31)$$
The contributions of this part can be neglected comparing with Eq.
(2.30) at high energy.

$$d\sigma_a^I(3)=\frac{Z^2\alpha^2}{4
E_i^2}\ln\frac{4E_i^2}{\mu^2}2\pi\frac{2E^2_i}{M_0^2}\int^{\pi}_{\theta_{min}}\frac{sin
\theta d\theta}{(\sin^2\frac{\theta}{2}+\mu^2/4E_i^2)^2}
\frac{sin^4\frac{\theta}{2}}
{(1+\frac{2E_i}{M_0}\sin^2\frac{\theta}{2})^2}
\frac{\alpha}{2\pi}\frac{[1+(1-z)^2](1-z)}{z}dz$$
$$=\frac{\pi Z^2\alpha^2}{2E_i^2}\ln\frac{4E_i^2}{\mu^2}\frac{E_i}{M_0}\int^{\pi}_{\theta_{min}}\frac{2sin
\frac{\theta}{2}cos\frac{\theta}{2}
d\theta}{(\sin^2\frac{\theta}{2}+\mu^2/4E_i^2)^2}
\frac{sin^4\frac{\theta}{2}}
{(\frac{M_0}{2E_i}+sin^2\frac{\theta}{2})^2}
\frac{\alpha}{2\pi}\frac{[1+(1-z)^2](1-z)}{z}dz$$
$$=\frac{\pi Z^2\alpha^2}{2E_i^2}\ln\frac{4E_i^2}{\mu^2}\frac{E_i}{M_0}\int^1_{\epsilon}
\frac{2dt}{(t+\mu^2/4E_i^2)^2} \frac{t^2}{(\frac{E_i}{E_0}+t)^2}
\frac{\alpha}{2\pi}\frac{[1+(1-z)^2](1-z)}{z}dz$$
$$=\frac{\pi
Z^2\alpha^2}{2E_i^2}\ln\frac{4E_i^2}{\mu^2}\frac{E_i}{M_0}\left[
\frac{1}{\frac{\mu^2}{4E_i^2}-\frac{M_0}{2E_i}}
\left(\frac{\frac{\mu^4}{16E^4_i}}{1+\frac{\mu^2}{4E_i^2}}
+\frac{\frac{M_0^2}{4E_i^2}}{1+\frac{M_0}{2E_i}}
-\frac{\frac{\mu^4}{16E^4_i}}{\epsilon+\frac{\mu^2}{4E^2_i}}-\frac{\frac{M_0^2}{4E_i^2}}{\epsilon+\frac{M_0}{2E_i}
} \right)\right.$$
$$\left.
-\frac{2}{(\frac{\mu^2}{4E^2_i}-\frac{M_0}{2E_i})^3}\ln\left\vert\frac{\frac{M_0}{2E_i}+1}
{\frac{\mu^2}{4E^2_i}+1}\right \vert
-\frac{2}{(\frac{\mu^2}{4E^2_i}-\frac{M_0}{2E_i})^3}\ln\left\vert\frac{\frac{M_0}{2E_i}+\epsilon}
{\frac{\mu^2}{4E^2_i}+\epsilon}\right \vert\right]
\frac{\alpha}{2\pi}\frac{[1+(1-z)^2](1-z)}{z}dz $$
$$\simeq
\frac{4Z^2\alpha^3E^2_i}{M_0^4}\ln\frac{E_iM_0}{m^2_e}\ln\frac{4E^2_i}{\mu^2}
\frac{[1+(1-z)^2](1-z)}{z}dz.\eqno(2.32)$$ The contributions of this
part can also be neglected comparing with Eq. (2.30) at
$E_i<100~GeV$. We will show that this restriction is unnecessary in
thin ionization gas since $m_e$ is replaced by $\mu$ in Eq. (2.30).

     However, the second term is not so lucky. Through the
numeric computations we find that the contribution of this term is
almost $\sim \beta Z^2\alpha^2/\mu^2\ln(4E_i^2/\mu^2)$ and
$\beta\sim -0.5$ is acceptable. Thus we have

$$d\sigma_a^I\simeq \frac{Z^2\alpha^3}{2m_e^2}
\ln\frac{4E^2_i}{\mu^2}\frac{[1+(1-z)^2](1-z)}{z}dz. \eqno(2.33)$$

   Now we calculate the process in Fig.5d. Corresponding to Eq.
(2.24) we have

$$d\sigma_b^I=\frac{m_eM_0}{\sqrt{(p_iP_i)^2-m^2_eM^2_0}}\vert\overline{M_{\hat{l}'P_i\rightarrow
p_fP_f}}\vert^2\frac{M_0d^3\vec{P_f}}{(2\pi)^3E^P_f}\frac{m_ed^3\vec{p_f}}{(2\pi)^3E_f}
(2\pi)^4\delta^4(p_i+P_i-p_f-P_f-k)$$
$$\times\left(\frac{1}{2E_{\hat{l}'}}\right)^2\left(\frac{1}{E_{\hat{l}'}+\omega-E_i}\right)^2
\vert\overline{M_{p_i\rightarrow \hat{l}'k}}\vert^2 \frac{2\pi
d^3\vec{k}}{(2\pi)^3\omega}$$
$$=\frac{Z^2\alpha^2}{4 E_i^2}\frac{1}{
(\sin^2\frac{\theta}{2}+\mu^2/4E_f(E_i-\omega))^2}
\frac{1-\sin^2\frac{\theta}{2}-\frac{q^2}{2M^2_0}\sin^2\frac{\theta}{2}}
{1+\frac{2(E_i-\omega)}{M_0}\sin^2\frac{\theta}{2}}$$
$$\times\frac{\alpha}{2\pi}\frac{1+(1-z)^2}{z(1-z)}d\Omega dzd\ln \frac{-q^2}{\mu^2}$$
$$\simeq\frac{Z^2\alpha^2}{4 E_i^2}\frac{1}{
(\sin^2\frac{\theta}{2}+\mu^2/4E_f^2)^2}
\frac{1-\sin^2\frac{\theta}{2}+\frac{2E_f^2\sin^4\frac{\theta}{2}}{M^2_0(
1-\frac{2E_f}{\mu}\sin^2\frac{\theta}{2})}}
{1+\frac{2E_f}{M_0}\sin^2\frac{\theta}{2}}$$
$$\times\frac{\alpha}{2\pi}\frac{1+(1-z)^2}{z(1-z)}d\Omega dzd\ln
\frac{-q^2}{\mu^2}, \eqno(2.34)$$where we used

$$\left(\sin^2\frac{\theta}{2}+\frac{\mu^2}{4E_f(E_i-\omega)}\right)^2
=\left(\sin^2\frac{\theta}{2}+\frac{\mu^2}{4E_f^2(1-\frac{2E_f}{M_0}\sin^2\frac{\theta}{2})^{-1})}\right)^2$$
$$=\left((1-\frac{\mu^2}{2E_fM_0})\sin^2\frac{\theta}{2}+\frac{\mu^2}{4E_f^2}\right)^2
\simeq \left(\sin^2\frac{\theta}{2}+\frac{\mu}{4E^2_f}\right)^2.
\eqno(2.35)$$

    We denote

$$d\sigma\equiv d\hat{\sigma}\frac{\alpha}{2\pi} Pdzd\Omega.\eqno(2.36)$$

    Using the exchange $i\leftrightarrow f$ between Eqs. (2.24) and (2.34),
we have

$$\hat{\sigma}_b^I=\hat{\sigma}_a^I\left(\frac{E_f}{E_i}\right)^2\simeq \hat{\sigma}_a^I(1-z)^2.
\eqno(2.37)$$On the other hand,

$$P_b=
\frac{1+(1-z)^2}{z(1-z)}.\eqno(2.38)$$Moving factor $(1-z)^2$ from
$\hat{\sigma}_b^I$ to $P_b$, we have

$$d\sigma_a^I=d\sigma_b^I,~~\hat{\sigma}_a^I=\hat{\sigma}_b^I,~~P_a=P_b,
\eqno(2.39)$$Thus, we have

$$d\sigma^I=d\sigma^I_a+d\sigma^I_b=\frac{Z^2\alpha^3}{m_e^2}
\ln\frac{4E^2_i}{\mu^2}\frac{[1+(1-z)^2](1-z)}{z}dz. \eqno(2.40)$$

    We keep leading order term $1/z$, i.e., the leading logarithmic
$(1/z)$ approximation ($LL(1/z)A$) and obtain the integrated cross
section

$$\sigma^I\simeq \frac{2Z^2\alpha^3}{m_e^2}\ln\frac{4E^2_i}{\mu^2}
\ln\frac{E_i}{\omega_{min}}.  \eqno(2.41)$$

We find that the parameter $m_e$ enters
$1/(\mu^2+m^2_e)\simeq 1/m^2_e$ through the lower limit of
integration. The resulting integrated cross section is
always suppressed by $\sim 1/m^2_e$ as similar to the BH formula
(see the next section).

We consider a different example: the electric field of ionized atom
can extend to a large space in the thin ionosphere, where the recoil
of the target atom can be neglected $\nu \ll E_i$ if the impact
parameter is large enough. Using $\delta(E_i-E_f-\omega)$ the
scattering potential becomes time-independent. Now Eq.(2.28) is
replaced by

$$\vert\vec{q}\vert^2=\vert\vec{p}_f-\vec{p}_i\vert^2
=4\vert\vec{p}\vert^2\sin^2\frac{\theta}{2}\simeq
4E_i^2\sin^2\frac{\theta}{2},\eqno(2.42)$$ where
$\vert{p}\vert\simeq E_i$ in the unit $c=1$. Corresponding to
Eq.(2.30) we have

$$d\sigma_a^{II}(1)$$
$$=\frac{\pi
Z^2\alpha^2}{2E_i^2}\ln\frac{4E_i^2}{\mu^2}\frac{M_0}{E_i}
\frac{1}{\frac{\mu^2}{4E_i^2}-\frac{M_0}{2E_i}}
\left(\frac{1}{1+\frac{\mu^2}{4E_i^2}}
+\frac{1}{\frac{M_0}{2E_i}-\frac{\mu^2}{4E^2_i}}\ln\frac{1+\frac{\mu^2}{4E^2_i}}{1+\frac{M_0}{2E_i}}
-\frac{1}{\epsilon+\frac{\mu^2}{4E^2_i}}-\frac{1}{\frac{M_0}{2E_i}+\frac{\mu^2}{4E^2_i}}
\ln\frac{\epsilon+\frac{\mu^2}{4E^2_i}}{\epsilon+\frac{M_0}{2E_i}}\right)$$
$$\times\frac{\alpha}{2\pi}\frac{[1+(1-z)^2](1-z)}{z}dz$$
$$\simeq
\frac{2Z^2\alpha^3}{\mu^2}\ln\frac{4E^2_i}{\mu^2}\frac{[1+(1-z)^2](1-z)}{z}dz,\eqno(2.43)$$where
$\epsilon=t_{min}=\mu^2/(4E^2_i)$, since the lower limit
$\theta_{min}$ of integral is determined by

$$\ln\frac{4E^2_i\sin^2\frac{\theta}{2}}{\mu^2}\geq 0 \eqno(2.44)$$

    Similar to Eqs. (2.40) and (2.41), we have

$$d\sigma^{II}=d\sigma^{II}_a+d\sigma^{II}_b=\frac{2Z^2\alpha^3}{\mu^2}
\ln\frac{4E^2_i}{\mu^2}\frac{[1+(1-z)^2](1-z)}{z}dz,
\eqno(2.45)$$and

$$\sigma^{II}\simeq \frac{4Z^2\alpha^3}{\mu^2}\ln\frac{4E^2_i}{\mu^2}
\ln\frac{E_i}{\omega_{min}}.  \eqno(2.46)$$

    The improved BH formula contains the factor $\ln 4E_i^2/\mu^2$
rather than $\ln 4E_i^2/m_e^2$ since we assume that all
bremsstrahlung photons have a transverse cut-off of order $\mu$ and
$\mu\ll m_e$. A main distinguish of two bremsstrahlung formulas is a
large difference in the cross section, which originates from
$1/\mu^2$ and $1/m^2_e$ in front of the logarithmic factor.
Comparing with this effect, the logarithmic difference between $\ln
1/\mu$ and $\ln 1/m_e$ can be neglected.

    The results show that the bremsstrahlung cross
section is almost proportional to the geometric area of the atomic
Coulomb field $\sim R^2$, rather than a weaker $\ln R$-dependence
that the BH formula predicted.

     The new bremsstrahlung formulas $d\sigma^{II}$ should include the
contributions of Fig. 3b,c since $\nu\sim 0$ . However, we will
prove that these corrections are negligible at high energy at Sec.
3.

\newpage
\begin{center}
\section{\bf Comparing with the Bethe-Heitler formula}
\end{center}

    The differential cross section of the BH formula [1] for bremsstrahlung
is

$$d\sigma_{B-H}
=\frac{Z^2\alpha^3}{{2\pi}^2}\frac{\vert\vec{p}_f\vert}{\vert\vec{p}_i\vert}\frac{d\omega}
{\omega}\frac{d\Omega_cd\Omega_k}{\vert
\vec{q}^4\vert}\left[\frac{\vec{p_f}^2\sin^2\theta_f}{(E_f-\vert\vec{p}_f\vert\cos
\theta_f)^2}(4E_i^2-\vec{q}^2)+\frac{\vec{p_i}^2\sin^2\theta_i}{(E_i-\vert\vec{p}_i\vert\cos
\theta_i)^2}(4E_f^2-\vec{q}^2)\right.$$
$$\left.+2\omega^2\frac{\vec{p}_i^2\sin \theta_i+\vec{p}_f^2\sin
\theta_f}{(E_i-\vert\vec{p}_i\vert\cos
\theta_i)(E_f-\vert\vec{p}_f\vert\cos \theta_f)}-2
\frac{\vert\vec{p_i}\vert\vert\vec{p_f}\vert\sin \theta_i\sin
\theta_f\cos \phi}{(E_i-\vert\vec{p}_i\vert\cos
\theta_f)(E_f-\vert\vec{p}_f\vert\cos
\theta_f)}(2E_i^2+2E_f^2-\vec{q}^2)\right], \eqno(3.1)$$ where
$\theta_i$, $\theta_f$ are the angles between $\vec{k}$ and
$\vec{p}_i$, $\vec{p}_f$ respectively, $\phi$ is the angle between
$(\vec{p_i}\vec{k})$ plane and $(\vec{p}_f\vec{k})$ plane.

    After integral over angles in Eq. (3.1) at $E\gg m_e$
(but still keeping $m_e$), the cross section can be simplified as

$$d\sigma_{B-H}\simeq
\frac{Z^2\alpha^3}{m_e^2}\frac{d\omega}{\omega} \frac{4}{E^2_i}
(E^2_i+E_f^2-\frac{2}{3}E_iE_f)\left(\log\frac{2E_iE_f}{m_e\omega}-\frac{1}{2}\right).\eqno(3.2)$$

    Unfortunately, Eq.(3.2) does not present
the screening effect since it uses a pure Coulomb potential. Bethe
and Heitler consider that the term $2E_iE_f/m^2_e\omega$ in Eq.(3.2)
is $\sim \mu$. Thus, they write

$$d\sigma_{B-H}=
\frac{Z^2\alpha^3}{m_e^2}\frac{d\omega}{\omega}\frac{4}{E^2_i}
(E^2_f+E_i^2-\frac{2}{3}E_fE_i)\left(\log
\frac{m_e}{\mu}-\frac{1}{2}\right),\eqno(3.3)$$or

$$d\sigma_{B-H}=
\frac{Z^2\alpha^3}{m_e^2}\frac{d\omega}{\omega}\frac{4}{E^2_i}
(E^2_f+E_i^2-\frac{2}{3}E_fE_i)\left(\log
(137Z^{-1/3})-\frac{1}{2}\right),\eqno(3.4)$$where the Thomas-Fermi
model is used.

    For comparison, we rewrite Eq.(3.3) as

$$d\sigma_{B-H}=\hat{\sigma}_{B-H}\frac{\alpha}{2\pi}P_{B-H}dv,\eqno(3.5)$$where

$$\hat{\sigma}_{B-H}=\frac{8\pi Z^2\alpha^2}{m^2_e}\left(\log
\frac{m_e}{\mu}-\frac{1}{2}\right),\eqno(3.6)$$and

$$P_{B-H}dz=\frac{d\omega}{\omega}\frac{1}{E_i^2}
\left(E_i^2+E^2_f-\frac{2}{3}E_iE_f\right).\eqno(3.7)$$

    Using $z=\omega/E_i$ (note that $\nu=0$ in the BH formula) and $d\omega/\omega=dz/z$, we
have

$$P_{B-H}(z)=\left[\frac{1+(1-z)^2}{z}
-\frac{2(1-z)}{3z}\right].\eqno(3.8)$$ Thus, the BH formula (3.3)
becomes

$$d\sigma_{B-H}\simeq
\frac{4Z^2\alpha^3}{m^2_e}\left(\log\frac{m_e}{\mu}-\frac{1}{2}\right)\left[\frac{1+(1-z)^2}{z}
-\frac{2(1-z)}{3z}\right]dz. \eqno(3.9)$$

It is well known that at limit $\omega\rightarrow 0$ any process
leading to photon emission can be factorized [13]. A corresponding
factorized differential cross section at this approximation is
[6,10]

$$\frac{d\sigma_{Soft}}{d\Omega}
=\frac{d\sigma_{Ruth}}{d\Omega}\frac{2\alpha}{\pi}\ln\frac{E_i}
{\omega_{min}}\left\{
\begin{array}{ll}
\frac{4}{3}\beta^2\sin^2\frac{\theta}{2},
& {NR}\\\\
\ln \frac{\vert\vec{q}\vert^2}{m^2_e}-1. & {ER}\end{array}\right.
\eqno(3.10)$$This equation describes the cross section for a single
photon radiation at the elastic limit. However, the integrated cross
section integrates over all possible phase space, it does not only
includes the contributions of elastic scattering $(\vert k\vert=0)$,
but also inelastic scattering $(\vert k\vert\sim 0)$. The
probability of such soft photons is proportional to [12]

$$W\sim \int^{\vert\vec{q}\vert}_{\mu}\frac{dk}{k}.\eqno(3.11)$$We
insert Eq.(3.11) to Eq.(3.10) at the ER limit and get

$$d\sigma_{Soft}^{ER}\simeq\int d\Omega
\frac{Z^2\alpha^2}{4E_i^2
\left(\sin^2\frac{\theta}{2}+\frac{\mu^2}{4E_i^2}\right)^2}
\frac{2\alpha}{\pi}\ln\frac{E_i}
{\omega_{min}}\left\vert\ln\frac{\vert\vec{q}\vert^2}{m^2_e}\ln\frac{\vert\vec{q}\vert^2}{\mu^2}-
\frac{1}{2}\ln\frac{\vert\vec{q}\vert^2}{\mu^2}\right\vert.\eqno(3.12)$$The
first term in the absolute value symbol is known as the Sudakov
double logarithm [12].

    We consider bremsstrahlung of high energy electrons in the normal
medium, where the nuclear Coulomb field is restricted inside the
atomic scale. The larger the electron energy, the larger the energy
transfer due to a stronger Coulomb field. In this case, the recoil
of the target atom can not be neglected even in solid since the
bound target can't completely absorb a strong recoil at such high
energy $(\gg 1~GeV)$. According to $d\sigma^I$ the integrated cross
section will be suppressed by a factor $1/m^2_e$ as similar to
$d\sigma_{B-H}$. For further comparison, we use the Coloumb
potential without recoil to replace $d\hat{\sigma}^I$ in Eq. (2.24)
and take the $LL(1/z)$ approximation, the result

$$d\sigma^I\sim\frac{Z^2\alpha^3}{E_i^2}
\int^{\pi}_{\theta_{min}}\frac{\sin \theta
d\theta}{(\sin^2\frac{\theta}{2}+\frac{\mu^2}{4E_i^2})^2}\ln\frac{-q^2}{\mu^2}\frac{dz}{z}
\simeq \frac{4Z^2\alpha^3}{m^2_e}\ln\frac{m_e^2}{\mu^2}\frac{dz}{z}.
\eqno(3.13)$$is consistent with the BH formula (3.9) at the same
approximation. There is a difference in the coefficients since we
neglect the contributions of Figs. 5b and 5d.

    On the other hand, $d\sigma^{II}$ shows that the bremsstrahlung cross section
has a big enhance at the thin ionized gas. We remember that the
cross section $d\sigma^{II}$ is valid at $\nu \rightarrow 0$, where
the contributions of Figs. 4b,c should be included. However, the
double logarithm in Eq.(3.12) contributes the following factor

$$d\sigma\propto \frac {Z^2\alpha^3}{m^2_e}\left[ \ln \frac{m^2_e}{\mu^2}+
\ln^2\frac{4E^2_i}{m^2_e}+\ln\frac{4E^2_i}{m^2_e}\ln\frac{4E^2_i}{\mu^2}+......\right],\eqno(3.14)$$
Obviously, these corrections are negligible if comparing with
$d\sigma^{II}$. Therefore, we suggest that $d\sigma^{II}$ is an
valuable bremsstrahlung formula in the thin ionized gas.

    According to the above discussions, we conclude that $d\sigma^I$ (or $d\sigma_{B-H}$)
applies to bremsstrahlung in most media, but we should use
$d\sigma^{II}$ for the bremsstrahlung process in the thin ionized
gas at high energy. We suggest to test our prediction in the
ionosphere.

    A straightforward way may help us to understand the anomaly
bremsstrahlung. For the electron going through the bremsstrahlung,
the probability for it to radiate a photon can be estimated by

$$Prob.\sim \frac{\alpha}{p\cdot k}P_{e\rightarrow
e\gamma}(z),\eqno(3.15)$$where $p$ is the electron momentum and $k$
is the photon momentum and $z=k^0/(k^0 + p^0)$ is the energy
fraction carried by the electron. Note that

$$p\cdot k=\sqrt{\vert \vec{p}\vert^2-m^2_e}\vert
\vec{k}\vert-\vert\vec{p}\vert\vert\vec{k}\vert\cos\theta> \frac{
\vert\vec{k}\vert}{2\vert\vec{p}\vert}m^2_e+\vartheta(m^4_e).\eqno(3.16)
$$ Therefore the probability is proportional to $m^{-2}_e$. However,
when there is the coulomb potential, a single momentum kick due to
the Coulomb potential is $\vec{k}_c\sim \vartheta (1/R)\ll m_e$,
which which will modify the denominator to

$$p\cdot k\rightarrow (p+k+k_e)^2-m^2_e\sim \frac{
\vert\vec{k}\vert}{2\vert\vec{p}\vert}m^2_e-2\vec{k}_c\cdot\vec{p}-\vec{k}^2_c,
\eqno(3.17)$$ it is still of the order of $m^2_e$ for the energy
region where is not large enough. However for extremely large
$\vert\vec{p}\vert$, there is enough room in the phase space for the
two terms to cancel, which results in

$$p\cdot k\sim -\vec{k}^2_c\sim \mu^2\ll m^2_e,\eqno(3.18)$$
and hence the probability may be enhanced to be proportional to
$1/\mu^2$.

\newpage
\begin{center}
\section{\bf The anomalous effect in electron-positron pair creation}
\end{center}

    Our discussion on bremsstrahlung also applies to electron-positron pair
creation. We consider a high energy photon traversing the atomic
Coulomb field. This photon has a certain probability of transforming
itself into a pair of electron-positron. The TOPT describes pair creation in Fig. 6. The
contributions of the interferant processes between Figs. 1a and 1c
can be neglected at the the thin ionosphere since it is also
$1/m^2_e$-suppressed.

    The cross section of pair creation in Fig. 7a at the leading
order approximation reads [10]

$$d\sigma_{\gamma\rightarrow e^+}=\frac{2\pi M_0}{E_i^P\omega\vert v-c\vert}\vert\overline{M_{kP_i\rightarrow
p_+P_fp_-}}\vert^2(2\pi)^4\delta^4(k+P_i-p_+-P_f-p_-)$$
$$\times\frac{m_e d^3\vec{p_+}}{(2\pi)^3E_+}\frac{m_ed^3\vec{p_-}}{(2\pi)^3E_-}
\frac{M_0d^3\vec{P_f}}{(2\pi)^3E^P_f},\eqno(4.1)$$where the
screening photon propagator in the matrix takes Eq. (2.2). We take
the laboratory frame, where the target atom is at rest, but the
incident photon has a high energy. Therefore, $M_0/E_i^P\simeq1$ and
note that $\vert v-c\vert\simeq c=1$ in the nature unit.

\begin{figure}
    \begin{center}
        \includegraphics[width=0.5\textwidth]{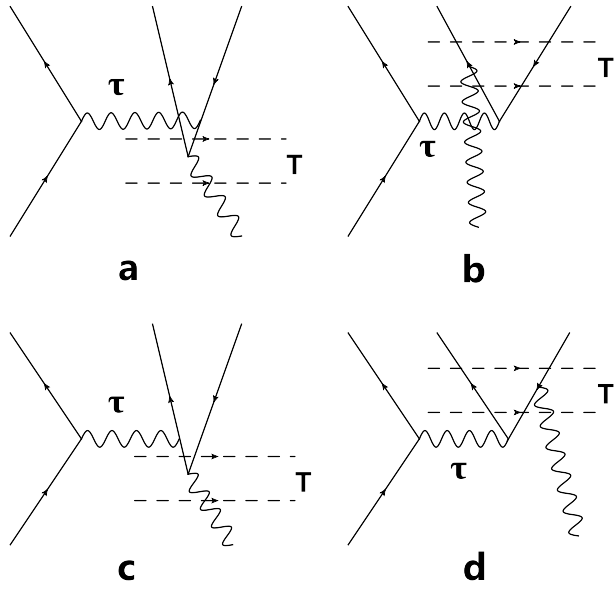}
        \caption{The TOPT decomposition for pair production.}\label{Fig6}
    \end{center}
\end{figure}

\begin{figure}
    \begin{center}
        \includegraphics[width=0.6\textwidth]{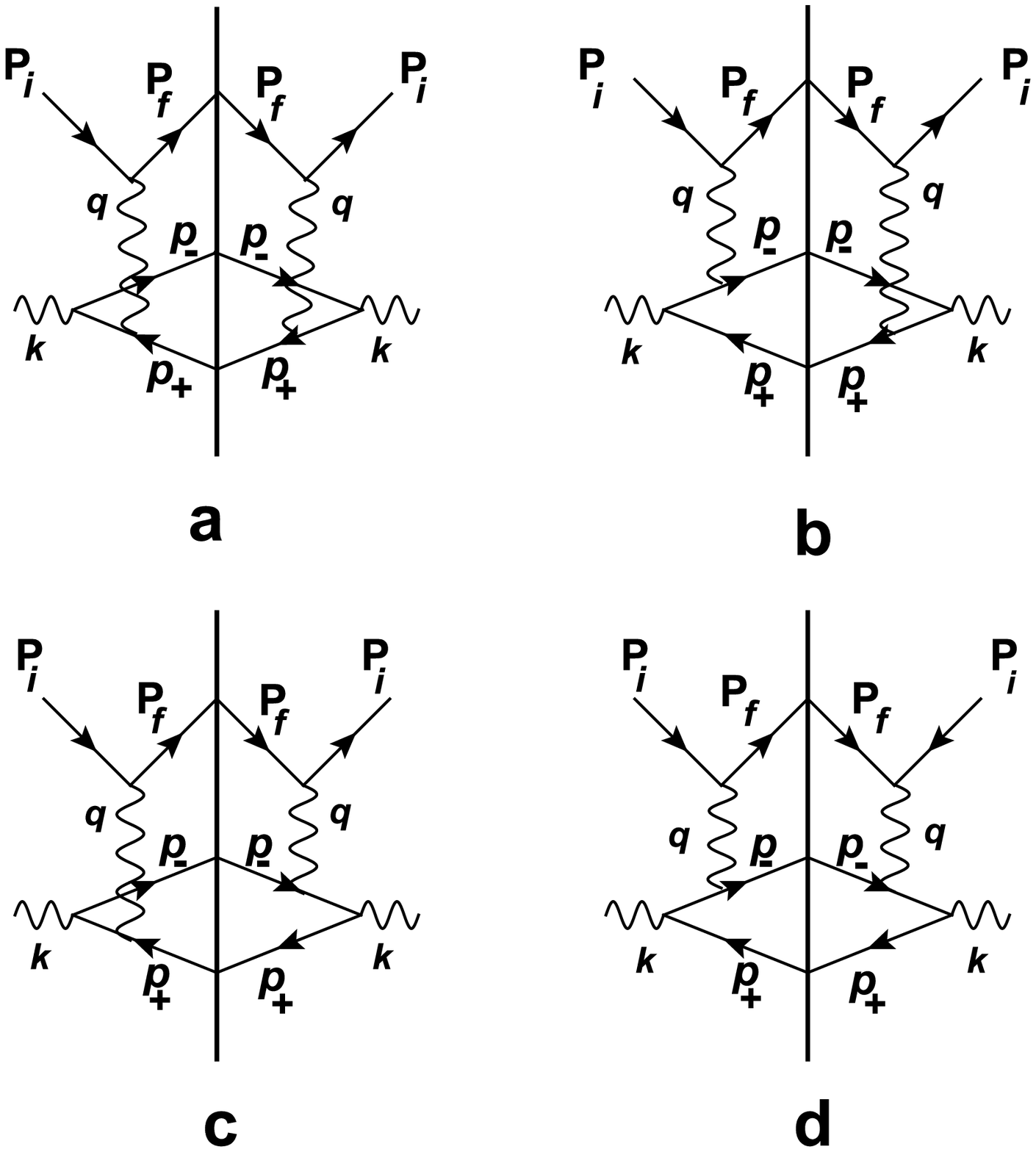}
        \caption{Four TOPT diagrams after neglecting the
            contributions of the backward components at high energy and small
            scattering angle.}\label{Fig7}
    \end{center}
\end{figure}

   The above cross section can be written as the factorization form in the TOPT framework,

$$d\sigma_{\gamma\rightarrow e^+}=
\left(\frac{1}{2E_{\hat{l}}}\right)^2\left(\frac{1}{E_{\hat{l}}+E_--\omega}\right)^2
\vert\overline{M_{k\rightarrow \hat{l}p_-}}\vert^2 \frac{
m_ed^3\vec{p}_-}{(2\pi)^2E_-\omega}$$
$$\times\vert\overline{M_{\hat{l}P_i\rightarrow
P_fp_+}}\vert^2\frac{M_0d^3\vec{P_f}}{(2\pi)^3E^P_f}\frac{m_ed^3\vec{p}_+}{(2\pi)^3E_+}
(2\pi)^4\delta^4(k+P_i-p_+-P_f-p_-).\eqno(4.2)$$where
$\hat{l}=(E_{\hat{l}}, \vec{l}_T, l_L)$ is on-mass shell, i.e.,
$\hat{l}^2=m^2_e$.

   We take $\vec{k}$ along the z-direction, and define $z$ as the momentum
fraction of $k=\vert \vec {k}\vert$ carried by the longitudinal
momentum of electron,

$$z=\frac{p_-^L}{k}\simeq
\frac{E_-}{\omega},\eqno(4.3)$$where we neglect the electron mass
$m_e$ at high energy. At high energy and small emitted angle we
denote $k=[\omega,0,\omega]$,
$p_-=[z\omega+\frac{l_T^2}{2z\omega},l_T,z\omega]$ and
$\hat{l}=[(1-z)\omega+\frac{l^2_T}{2(1-z)\omega}, -l_T,
(1-z)\omega]$. Trough a simple calculation, one can get

$$d\sigma_{\gamma\rightarrow e^-}\equiv d\hat{\sigma}d\Omega
\frac{\alpha}{2\pi} Pdz\ln \vec{l}^2_T$$
$$=\vert\overline{M_{\hat{l}P_i\rightarrow
P_fp_+}}\vert^2\frac{2M_0}{(2\pi)^2}d^4P_f\delta(P^2_f-M^2_0)\Theta(E_f^P)m_e\vert
\vec{p}_-\vert dE_-d\Omega\frac{\alpha}{2\pi}Pdzd\ln \vec{l}^2_T$$
$$=\frac{2m_eM_0}{(2\pi)^2}\vert \vec {p}_-\vert \frac{\vert\overline{M_{\hat{l}P_i\rightarrow
P_fp_+}}\vert^2}{2[M_0+(\omega-E_-)-(\omega-E_-)\cos
\theta]}d\Omega\frac{\alpha}{2\pi}Pdzd\ln \vec{l}^2_T$$
$$=\frac{Z^2
\alpha^2}{4\omega
m_e}\frac{1}{\left(\sin^2\frac{\theta}{2}+\frac{\mu^2}{4E_+^2}\right)^2}
\frac{\cos^2\frac{\theta}{2}-\frac
{q^2}{2M_0^2}\sin^2\frac{\theta}{2}}{1+\frac{2(\omega-E_-)}{M_0}\sin^2\frac{\theta}{2}}d\Omega
\frac{\alpha}{2\pi}Pdzd\ln \vec{l}^2_T$$
$$=\frac{Z^2
\alpha^2}{4\omega^2}\frac{1}{\left(\sin^2\frac{\theta}{2}+\frac{\mu^2}{4E_+^2}\right)^2}
\frac{\cos^2\frac{\theta}{2}-\frac
{q^2}{2M_0^2}\sin^2\frac{\theta}{2}}{1+\frac{2(\omega-E_-)}{M_0}\sin^2\frac{\theta}{2}}d\Omega$$
$$\times
\frac{\alpha}{2\pi}z\left(\frac{1-z}{z}+\frac{z}{1-z}\right)dzd\ln
\vec{l}^2_T.\eqno(4.4)$$At the last step, we use

$$\frac{1}{2E_{\hat{l}}}\simeq\frac{1}{2(1-z)\omega},\eqno(4.5)$$

$$\frac{1}{E_{\hat{l}}+E_--\omega}\simeq\frac{1}{(1-z)\omega+\frac{l^2_T}{2(1-z)
\omega}+z\omega+\frac{l^2_T}{2z\omega}-\omega}
=\frac{2z(1-z)\omega}{l^2_T},\eqno(4.6)$$

$$\vert\overline{M_{k\rightarrow \hat{l}p_-}}\vert^2=2\alpha
l^2_T\left(\frac{1-z}{z}+\frac{z}{1-z}\right), \eqno(4.7)$$to
calculate

$$\frac{\alpha}{2\pi}Pdz\ln
\vec{l}^2_T=
\frac{2\pi}{\omega}\left(\frac{1}{2E_{\hat{l}}}\right)^2
\left(\frac{1}{E_{\hat{l}}+E_--\omega}\right)^2
\vert\overline{M_{k\rightarrow \hat{l}p_-}}\vert^2 \frac{
m_ed^3\vec{p}_-}{(2\pi)^3E_-}$$
$$=\frac{\alpha}{2\pi}\frac{m_e}{\omega}z\left(\frac{1-z}{z}+\frac{z}{1-z}\right)dzd\ln \vec{l}^2_T, \eqno(4.8)$$
where $m_e/\omega$ will move to $d\hat{\sigma}$.

    Now we integral over angle in Eq. (4.4) and result is

$$d\sigma_{\gamma\rightarrow
e^+}=\frac{\alpha^3}{2\mu^2}\ln\frac{4\omega^2}{\mu^2}
(1-z)\left[(1-z)^2+z^2\right]dz. \eqno(4.9)$$
Considering$\gamma\rightarrow e^+=\gamma\rightarrow e^-$, we have

$$d\sigma_{\gamma\rightarrow e}=d\sigma_{\gamma\rightarrow e^+}+d\sigma_{\gamma\rightarrow
e^-}=\frac{\alpha^3}{\mu^2}\ln\frac{4\omega^2}{\mu^2}(1-z)
[(1-z)^2+z^2]dz, \eqno(4.10)$$where $e=e^++e^-$.

\newpage
\begin{center}
\section{The improved electromagnetic cascade equation}
\end{center}

    High-energy electrons traversing matter lose energy
by radiation. The secondary photon products a pair of
electron-positron, which can further radiate. At each step the
number of particles increases while the average energy decreases,
until the energy falls below the critical energy. This phenomenon is
called the electromagnetic shower. The evolution of the energy
spectra of electrons and photons in a shower is described by the
cascade equation, which couples bremsstrahlung of electron and pair
production of photon. The theoretical basis of the electromagnetic
cascade equation is the BH formula for bremsstrahlung and pair
creation. Therefore, we modify the cascade equation in this section.

   We denote $X$ and $\lambda$ as the depth and the
radiation length in unity $g/cm^2$. The photon flux $\Phi_{\gamma}$
and electron/positron flux $\Phi_e$ satisfy the coupled equations in
the electromagnetic cascade process [14]

$$\frac{d\Phi_{\gamma}(\omega,
t)}{dt}=\int^{\infty}_{\omega} \frac {dE_i}{E_i}P_{e\rightarrow
\gamma}\left(\frac{\omega}{E_i}\right)\Phi_e(E_i,t)-
\Phi_{\gamma}(\omega,t)\int_0^1 dzP_{\gamma\rightarrow e}(z),
\eqno(5.1)$$and

$$\frac{d\Phi_e(E_f,
t)}{dt}=\int^{\infty}_{E_f} \frac {dE_i}{E_i}P_{e\rightarrow
e}\left(\frac{E_f}{E_i}\right)\Phi_e(E_i,t)- \Phi_e(E_f,t)\int_0^1
dzP_{e\rightarrow e}(z)+\int^{\infty}_{E_f} \frac
{d\omega}{\omega}P_{\gamma\rightarrow
e}\left(\frac{E_f}{\omega}\right)\Phi_{\gamma}(\omega,t).\eqno(5.2)$$Where
the cascade kernel $P_{e\rightarrow\gamma}(z)dtdz$ (or
$P_{e\rightarrow e}(z)dtdz$) is the probability for an
electron/positron of energy $E_i$ to radiate a photon of energy
$\omega=zE_i$ (or to an electron/positron of energy $E_f=zE_i$) in
traversing $dt=dX/\lambda$, while $P_{\gamma\rightarrow e}(z)dtdz$
is the probability for a photon of energy $\omega$ to radiate an
electron/positron of energy $E_f=zE_i$ in traversing
$dt=dX/\lambda$.

    A following key step is to extract the cascade kernels from
the bremsstrahlung- and pair production-cross creation. The
logarithmic $\ln 4E_i^2/\mu^2$ changes slowly with energy, it can be
regarded a constant. The cascade kernels are irrelevant to the
energy and they are functions of $z$ in a so-called approximation A
[14]. Interestingly, comparing with the QCD evolution equation [15],
we find that the corresponding kernels in Eqs. (2.45) and (4.10)
have similar form as that in the QCD equation except the different
normalization coefficients and two factors $E_f/E_i$ and
$E_-/\omega$. The former is because of the reason that Eqs. (5.1)
and (5.2) are scaled by the radiation length $\lambda$, while the
later two factors are arisen from the definition of the parton
(electron) distribution. We imitate the QCD evolution equation and
suggest to insert

$$1=\frac{E_f}{E_i}\frac{E_i}{E_f}, \eqno(5.3)$$and

$$1=\frac{E_-}{\omega}\frac{\omega}{E_-}, \eqno(5.4)$$into Eqs. (2.40)
and (4.10), where $E_f/E_i$ (or $E_-/\omega)$ incorporates with the
flux $\Phi_e$ (or $\Phi_{\gamma}$), while $E_i/E_f$ (or
$\omega/E_-$) belongs to the cascade kernels. Using the normalized
condition [14]

$$\int^1_0dz zP_{e\rightarrow e}(z)=1,\eqno(5.5)$$we get

$$P_{e\rightarrow e}=\frac{3}{4}\frac{1+z^2}{1-z}, \eqno(5.6)$$

$$P_{e\rightarrow
\gamma}=\frac{3}{4}\frac{1+(1-z)^2}{z},\eqno(5.7)$$and

$$P_{\gamma\rightarrow
e}=\frac{3}{4}[z^2+(1-z)^2].\eqno(5.8)$$ Our numeric solutions show
that the above mentioned substitutions don't influent the anomalous
bremsstrahlung effect.

    The traditional cascade equation has a same form as Eqs. (5.1) and (5.2) but with
the different cascade kernels, they are [14]

$$P_{e\rightarrow\gamma}(z)=z+\frac{1-z}{z}\left(\frac{4}{3}+2b\right), \eqno(5.9)$$

$$P_{e\rightarrow
e}(z)=1-z+\frac{z}{1-z}\left(\frac{4}{3}+2b\right)\eqno(5.10)$$and

$$P_{\gamma\rightarrow
e}(z)=\frac{2}{3}-\frac{1}{2}b+(\frac{4}{3}+2b)(z-\frac{1}{2})^2.
\eqno(5.11)$$The parameter $b=[18\ln(183/Z^{1/3})]^{-1}=0.0122$
relates with the screening effects.

\begin{figure}
    \begin{center}
        \includegraphics[width=0.6\textwidth]{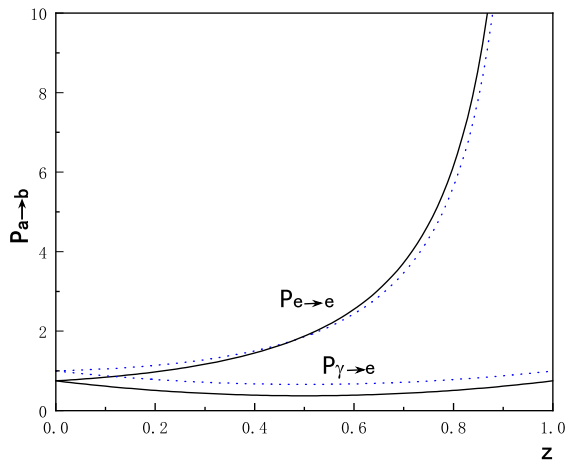}
        \caption{Comparison of the cascade kernels for a new cascade equation (the solid
    curves) and the BH-formula-based equation (the dashed curves)}\label{Fig8}
    \end{center}
\end{figure}

    We compare the cascade kernels of the traditional equation and
our improved equation in Fig. 8. One can find the difference between
them is negligible. Note that the cascade equation is scaled by the
radiation length $\lambda$. Therefore the integrated cross sections
$\sigma_{a\rightarrow b}$ are not appear in the cascade equation.
The solution $\Phi(E,t)$ should be replaced by $\Phi(E,X)$ using
$X=t\lambda$, where $\lambda$ contains the anomalous effect in
bremsstrahlung and pair creation.

   Note that the cascade kernel is normalized in Eq. (5.5),
therefore, the anomalous bremsstrahlung effect does not directly
appear in the solutions of the cascade equation. When we estimate
the value of the radiation length $\lambda$ as shown in Sec. 1, this
effect is shown by the cross section.

\newpage
\begin{center}
\section{\bf Summary}
\end{center}

    The BH formula successfully describes bremsstrahlung of high
energy electrons. The integrated cross section of the BH formula is
restricted in a region $\sim 1/m^2_e$, which is much smaller than
the geometric section of the target. Recently, the measured energy
spectra of electrons-positrons at the GeV-TeV energy band in cosmic
rays show two different sets. According to the traditional
bremsstrahlung theory, the above difference can't be caused by the
the electromagnetic shower at the top of atmosphere, since the small
integrated cross section implies that the energy loss of the shower
at the thin ionosphere is negligible.

    We find that an anomalous effect in bremsstrahlung and pair
creation arises an unexpected big increment at the atmosphere top,
which is missed by previous theory. This anomalous effect is caused
by the accumulation of a large amount of soft photon radiation. We
derive the relating formula including an improved electromagnetic
cascade equation. These results may use to explain the above
confusion in the electron-positron spectra.

\noindent {\bf ACKNOWLEDGMENTS} I would like to thank an anonymous
reviewer for his/her insightful comments on the manuscript.
Especially, the statement in Eqs. (3.15)-(3.18) is proposed by this
reviewer. The work is supported by the National Natural Science of
China (No.11851303).

\end{document}